\documentclass[10pt,a4paper]{article}
\usepackage{jheppub_kim}
\usepackage{pdflscape}
\usepackage{amsmath}
\usepackage{amssymb}
\usepackage{dcolumn}
\usepackage{bm}
\usepackage{color}
\usepackage{epsfig}
\usepackage{amsfonts}
\usepackage{graphicx}
\usepackage{subfigure}
\usepackage{dcolumn}
\usepackage{color}

\begin{document}
\title{Two-Field Inflationary Model and Swampland de Sitter Conjecture}

\author[a]{S. Noori Gashti}

\affiliation[a] {Department of Physics, Faculty of Basic Sciences, University of Mazandaran P. O. Box 47416-95447, Babolsar, Iran.}

\emailAdd{saeed.noorigashti@stu.umz.ac.ir}

\abstract{In this paper, we are going to investigate a new perspective of the two-field inflation model with respect to the swampland dS conjecture. At the first step, we study the two-fields inflation model, and apply the swampland conjecture to our model. Then, we calculate some cosmological parameters such as scalar spectrum index, tensor-to-scalar ratio, and compare our results with the recent observational data. Also, we give numerical analysis to show agreement with observational data.}


\maketitle

\section{Introduction}
Recently, cosmological theories and observations have provided interesting information about the universe. Data and measurements from the cosmic microwave background (CMB) show that the matter and energy fluctuations are always unstable on a large scale \cite{2,3}. The leading cause of these cosmic fluctuations is still unknown, and there is no specific explanation. But, like all other phenomena, cosmologists use different scenarios to explain the reasons for these fluctuations, including the inflationary world \cite{6,7,8}. Hence, cosmic inflation is a model for the production of perturbations related to the initial density of the universe, which somehow involves the structure formation. As can be seen from the inflation patterns, the universe has gone through an early period of accelerated expansion to solve the problems in cosmology, such as the horizon, flatness, and monopole problems \cite{10}. This accelerated expansion also led to these quantum fluctuations, and over time, these perturbations intensified under gravity, creating the structure of galaxies and everything in the universe on a large scale \cite{10,12,14}. Researchers have studied different inflation models that the simplest model is as inflation by a slow-roll scalar field \cite{B1, B2}. However, many reasons show that inflationary models may be practical for more than one field. First, in many theories, such as string theory or supersymmetry, and many other areas, we are practically dealing with several fields. Second, using two or more scalar fields, may offer desirable features and have many implications in cosmology. For example, hybrid models that include two scalar fields achieve both inflationary ranges and the area of density fluctuations. These are consistent with observable data and they occur at sub-Planckian scales\cite{17,18,19,20}. However, due to the advantages of studying multi-field inflation models, their analysis is complicated and has particular complexities due to the observable data. When we are faced with two or more scalar fields, perturbations in the relative contribution to the energy density are also possible, along with perturbations in total energy density \cite{22}. These isotropic perturbations may be source of curvature perturbations, and their evolution at the super-horizon scale, which confronts calculations with certain complexities, such as the density power spectrum \cite{26,27,28}. As a result, these fields are associated with several initial conditions that affect the power spectrum \cite{29,31}. The important thing is that the complexity of these multi-field inflation models will show against the observable data. Therefore, we should always consider a complete framework for these models and test multi-field models \cite{31,32,33,38,39,41,42,43}. One of these cases is the use of two-field inflation models. In this article, we want to analyze a new perspective of these two-field inflation models according to specific conditions. For two-field inflation, a number of specific models have been considered already \cite{46,47,48,49,52,53,54,55,56,57,59,60,61,62,63,64,65,66,67}. In general, two-field inflation has been used in many cases and has several implications in cosmology. It is including approximate solutions to the metric perturbations in slow-roll approximation. Also, this scenario include the evolution equation for adiabatic and entropy perturbations for certain models with kinetic corrections. There are also models with specific non-canonical corrections, or some unconventional kinetic corrections \cite{68,70,71,72,73}.\\
Given all the concepts mentioned above, we now want to consider a two-field inflation model concerning swampland conjecture \cite{80}. A conjecture, called weak gravity, has recently been introduced \cite{76,77,78,79,q,w}. According to this conjecture, gravity introduced as the weakest force at high energy limit of theories coupled to the gravity \cite{81,83,84}. There is an area that is consistent with quantum gravity called landscape, but at low energy, the landscape is surrounded by a larger area called swampland that contradicts quantum gravity \cite{85,86,86b}.\\
For inflation models to be compatible with quantum gravity, they must meet two criteria. Some inflation models consistent with these criteria, and many inflation models were inconsistent with them. In general, these two conditions are called swampland distance conjecture, which provides an upper limit for $\Delta\phi$ (variation of scalar fields), and swampland de Sitter (dS) conjecture, which provides a limit for potential slope. In this article, we use the swampland dS conjecture. In general, based on Planck mass $M_{pl}$, these conditions are expressed in the following forms \cite{88,89,90,91,e},
\begin{equation}\label{1}
\frac{\Delta\phi}{M_{pl}}<\mathcal{O}(1),
\end{equation}
and
\begin{equation}\label{2}
M_{pl}\frac{V'}{V}>c,
\end{equation}
where $V$ is a scalar field potential, and $c$ is a positive constant.\\
Given the above concepts, in this paper, we present a two-field inflation model method based on the swampland dS conjecture. Then, we evaluate the compatibility or incompatibility of this model with respect to observable data. This paper is organized as follows. In section 2, we study the two-field inflation model and introduce different types of cosmological parameters. In section 3, according to the swampland dS conjecture and the concepts expressed in this paper, we reobtain all of the parameters in section 2 with new points of view. Then, we investigate the compatibility or incompatibility of this inflation model according to the observable data by plotting some figures related to each of these cosmological parameters. We determine the range of these parameters in the final section before conclusion.

\section{Two-field inflation model}
In this section, we first briefly introduce a two-field inflation model, then evaluate these expressed models with the swampland dS conjecture. In order to examine the inflation models, we first consider the corresponding action and metrics. So, for two scalar fields, the action has following form,
\begin{equation}\label{3}
S=\int d^{4}x\sqrt{-g}\left(\frac{M_{pl}^{2}}{2}R-\frac{1}{2}\partial_{\mu}\phi \partial^{\mu}\phi-\frac{1}{2}\partial_{\mu}\chi \partial^{\mu}\chi-V(\phi,\chi)\right),
\end{equation}
where $R$ and ($\phi,\chi$) are Ricci scalar and two scalar fields, respectively, and $V(\phi,\chi)$ is corresponding potential. The Friedmann-Robertson-Walker (FRW) space-time introduced as following \cite{91b},
\begin{equation}\label{4}
ds^{2}=-dt^{2}+\alpha^{2}(t)\delta_{ij}dx^{i}dx^{j}.
\end{equation}
Concerning the above equation and scale factor $\alpha(t)$, the Friedmann equations for the evolution of $\alpha(t)$ are as follows \cite{92},
\begin{equation}\label{5}
H^{2}=\frac{1}{3M_{pl}^{2}}(\frac{1}{2}\dot{\phi}^{2}+\frac{1}{2}\dot{\chi}^{2}+V(\phi,\chi)),
\end{equation}
and
\begin{equation}\label{6}
-2\dot{H}=\frac{1}{M_{pl}}^{2}(\dot{\phi}^{2}+\dot{\chi}^{2})
\end{equation}
The slow-roll parameters such as $\epsilon$ and $\eta$ are given by,
\begin{equation}\label{7}
\epsilon=\frac{3(\dot{\phi}^{2}+\dot{\chi}^{2})}{\dot{\phi}^{2}+\dot{\chi}^{2}+2V},
\end{equation}
and
\begin{equation}\label{8}
\eta=-\frac{2(\dot{\phi}\ddot{\phi}+\dot{\chi}\ddot{\chi})}{H(\dot{\phi}^{2}+\dot{\chi}^{2})}.
\end{equation}
According to the above concepts and using the Hamilton-Jacobi equation, and also the description of inflation dynamics, we consider the following form of Hubble parameter \cite{92},
\begin{equation}\label{9}
H=H_{0}+H_{1}\phi+H_{2}\chi.
\end{equation}
Also, equations of motion with respect to the mentioned method is as follows,
\begin{equation}\label{10}
\dot{\phi}=-2M_{pl}^{2}\frac{dH}{d\phi},
\end{equation}
and
\begin{equation}\label{11}
\dot{\chi}=-2M_{pl}^{2}\frac{dH}{d\chi}
\end{equation}
Solving differential equations (\ref{10}) and (\ref{11}) one can obtain,
\begin{equation}\label{12}
\phi(t)=-2M_{pl}^{2}H_{1}t+\phi_{0}
\end{equation}
and
\begin{equation}\label{13}
\chi(t)=-2M_{pl}^{2}H_{2}t+\chi_{0}
\end{equation}
Also, the corresponding potential of two scalar fields calculated as following,
\begin{eqnarray}\label{14}
V(\phi,\chi)&=&(3H_{0}^{2}-2M_{pl}^{2}H_{1}^{2}-2M_{pl}^{2}H_{2}^{2})+6H_{1}H_{2}\phi\chi\nonumber\\
&+&6H_{0}H_{1}\phi+6H_{0}H_{2}\chi+3H_{1}^{2}\phi^{2}+3H_{2}^{2}\chi^{2}.
\end{eqnarray}
In order to calculate the scalar spectral index $(n_{s})$ and tensor-to scalar ratio \cite{EPJP}, we can use the power spectrum with respect to $C_{s}k=\alpha H$ with $C_{s}^{2}=1$ and $W_{s}\equiv\frac{(\dot{\phi}^{2}+\dot{\chi}^{2})}{2M_{pl}^{2}H^{2}}$,  which yields,
\begin{equation}\label{15}
A_{s}=\frac{H^{2}}{8\pi^{2}W_{s}C{s}^{2}}.
\end{equation}
Therefore, we can obtain,
\begin{equation}\label{16}
n_{s}-1=-2\epsilon-\frac{1}{H}\frac{d}{dt}\ln\epsilon,
\end{equation}
and
\begin{equation}\label{17}
r=16\epsilon.
\end{equation}
We have introduced a two-field inflation model, and we expressed different values for cosmological parameters. In addition to the above, other cosmological parameters such as the number of enfolds, running spectrum index, etc., can be obtained. In next section, we reproduce all these values related to the above concepts. We investigate the two-fields inflation modes with respect to swampland dS conjecture. Finally, we plot some figures to determine the ranges of each of these parameters.

\section{The de Sitter conjecture}
According to all mentioned motivations, and by using a series of direct calculations, we will recalculate the potential and other cosmological parameters. We also specify the range associated with each of the parameters using the swampland dS conjecture. Then, we plot some figures, and determine these ranges. Hence, concerning equations of previous section, the potential can be written as following,
\begin{equation}\label{18}
V=3(H_{0}-2M_{pl}^{2}(H_{1}^{2}+H_{2}^{2})t)^{2}-2M_{pl}^{2}(H_{1}^{2}+H_{2}^{2}).
\end{equation}
Now, according to equation (\ref{2}) and above equation, we obtain,
\begin{equation}\label{19}
\frac{6H_{1}H_{2}}{3(H_{0}-2M_{pl}^{2}(H_{1}^{2}+H_{2}^{2})t)^{2}-2M_{pl}^{2}(H_{1}^{2}+H_{2}^{2})}>c.
\end{equation}
The above equation is the dS conjecture concerning mentioned equations. Also, using the equation (\ref{5}) we can write,
\begin{equation}\label{20}
\epsilon=\frac{4M_{pl}^{2}(H_{1}^{2}+H_{2}^{2})}{2M_{pl}^{2}(H_{0}-2M_{pl}^{2}(H_{1}^{2}+H_{2}^{2})t)^{2}}.
\end{equation}
The scalar spectral index $n_{s}$ and tensor-to-scalar ratio ($r$) regarding the equations (\ref{16}) and (\ref{17}), one can obtain,
\begin{equation}\label{21}
n_{s}-1=-\frac{8M_{pl}^{2}(H_{1}^{2}+H_{2}^{2})}{(H_{0}-2M_{pl}^{2}(H_{1}^{2}+H_{2}^{2})t)^{2}},
\end{equation}
and
\begin{equation}\label{22}
r=16\frac{4M_{pl}^{2}(H_{1}^{2}+H_{2}^{2})}{2M_{pl}^{2}(H_{0}-2M_{pl}^{2}(H_{1}^{2}+H_{2}^{2})t)^{2}}.
\end{equation}
After calculating the above-mentioned values related to cosmological parameters, we will examine the relations proportional to the swampland dS conjecture. It is noteworthy that by inverting equation (\ref{21}), two values are obtained according to the scalar spectrum index, and we will have two relations by placing them in the equation (\ref{19}). Therefore, according to the equation (\ref{21}), by inverting the function and obtaining the relation according to the scalar spectral index $(n_{s})$, and replacing it in the equation (\ref{19}), the equations related to swampland conjecture convert to the following forms,
\begin{equation}\label{23}
-\frac{6M_{pl}^{2}(-1+n_{s})^{2}t^{2}H_{1}H_{2}}{(11+n_{s})(M_{pl}^{2}(-1+n_{s})tH_{0}-2(M_{pl}^{2}+\sqrt{M_{pl}^{4}(1-(-1+n_{s})tH_{0})}))}>c,
\end{equation}
and
\begin{equation}\label{24}
-\frac{6M_{pl}^{2}(-1+n_{s})^{2}t^{2}H_{1}H_{2}}{(11+n_{s})(M_{pl}^{2}(-2+(-1+n_{s})tH_{0})+2\sqrt{M_{pl}^{4}(1+(t-n_{s}t)H_{0})})}>c.
\end{equation}
Now, we perform the same procedure for another parameter, i.e., the tensor-to-scalar ratio. So, according to the equation (\ref{22}), by inverting the function and obtaining the relation in terms of tensor-to-scalar ratio ($r$) and replacement in the equation (\ref{19}), the equations related to swampland conjecture are given as follows. In this part, precisely like the previous part, two different values are obtained,
\begin{equation}\label{25}
-\frac{6(4\sqrt{M_{pl}^{4}(4+rtH_{0})}+M_{pl}^{2}(8+rtH_{0}))H_{1}H_{2}}{M_{pl}^{2}(-48+r)H_{0}^{2}}>c,
\end{equation}
and
\begin{equation}\label{26}
-\frac{6M_{pl}^{2}r^{2}t^{2}H_{1}H_{2}}{(-48+r)(4\sqrt{M_{pl}^{4}(4+rtH_{0})}+M_{pl}^{2}(8+rtH_{0}))}>c.
\end{equation}
After the above calculations, we obtain the range appropriate to each cosmological parameters. Therefore,  we determine the range of each of these cosmological parameters by plotting some figures. Of course, with respect to the values of (\ref{16}) and (\ref{17}), the relation between the two parameters of cosmology can be well obtained. Now, according to the above concepts, we plot some figures related to each cosmological parameter, so  as you can see in the figures, we have plotted the range of each of the cosmological parameters, such as the scalar spectrum index and the tensor-to-scalar ratio, according to the swampland conditions. The area of each parameter has been compared according to the observable data.\\
Assuming $M_{pl}=1$, we can obtain behavior of the swampland dS conjecture  using the equation (\ref{23}) which is represented by Fig. \ref{1}. For $n_{s}<1$, it is decreasing function, while for $n_{s}>1$ it is increasing function, it vanishes for $n_{s}=1$. Also, we show that increasing Hubble parameter increases value of $c$ parameter. In Fig. \ref{1} (a) we vary $H_{2}$, while in Fig. \ref{1} (b) vary $H_{1}$ to see similar result. In Fig. \ref{1} (c) we vary $H_{0}$ and see that is not so important parameter.

\begin{figure}[h!]
 \begin{center}
 \subfigure[]{
 \includegraphics[height=4cm,width=4cm]{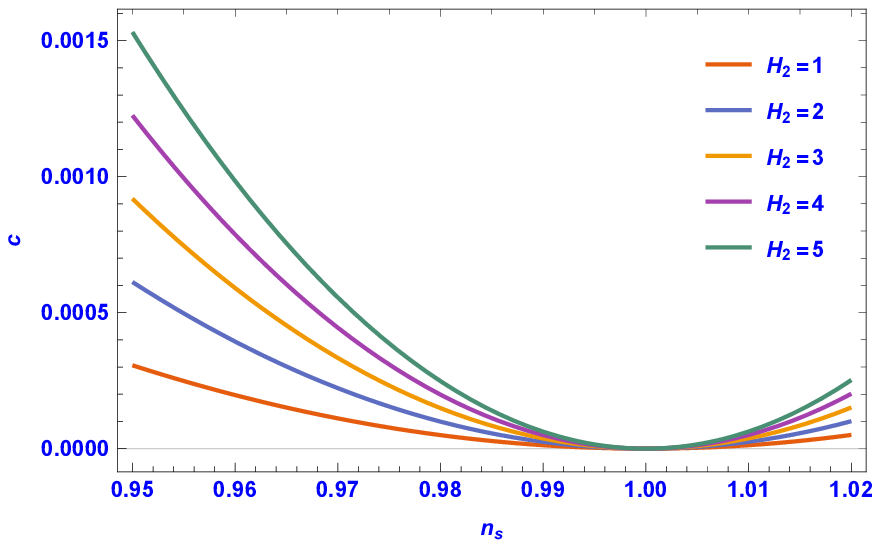}
 \label{1a}}
 \subfigure[]{
 \includegraphics[height=4cm,width=4cm]{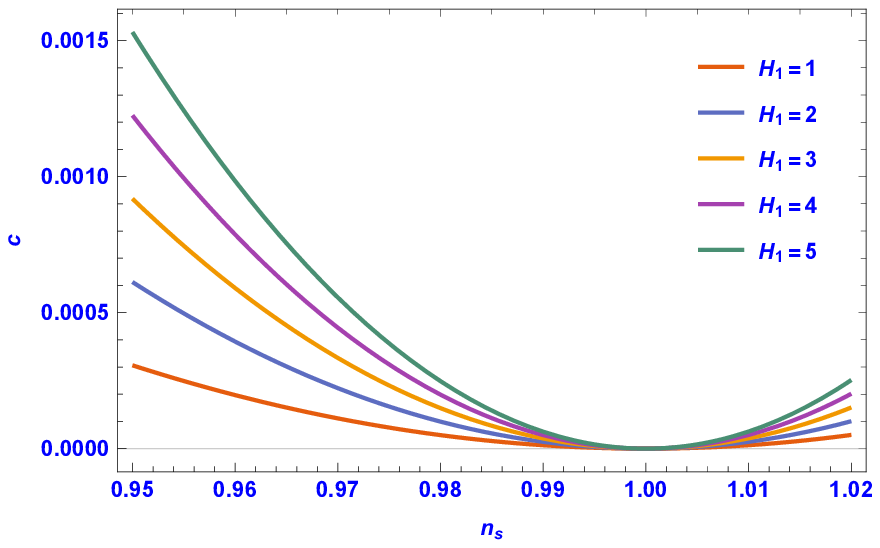}
 \label{1b}}
 \subfigure[]{
 \includegraphics[height=4cm,width=4cm]{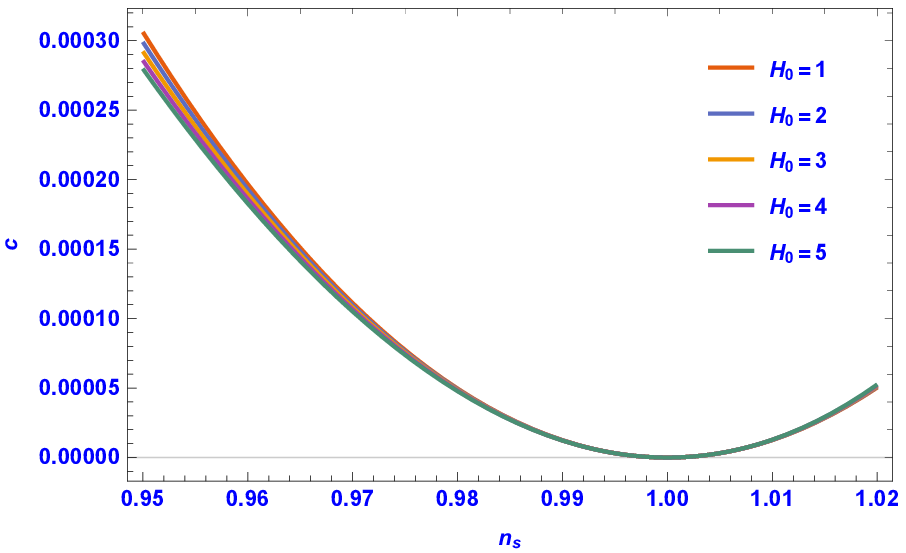}
 \label{1c}}
  \caption{\small{Behavior of the swampland dS conjecture ($c$) in term of $n_{s}$ in units of constant parameters, consistent to the Eq. (\ref{23}).}}
 \label{1}
 \end{center}
 \end{figure}

Similarly, we obtained behavior of the swampland dS conjecture in term of $n_{s}$ consistent to the equation (\ref{24}) and see that is decreasing function of $n_{s}$ with slow variation (see Fig. \ref{2}). Results are symmetric by variation of $H_{2}$ (Fig. \ref{2} (a)) or $H_{1}$ (Fig. \ref{2} (b)). In Fig. \ref{2} (c) we assumed unit values for $H_{2}$ and $H_{1}$, and vary $H_{0}$.

\begin{figure}[h!]
 \begin{center}
 \subfigure[]{
 \includegraphics[height=4cm,width=4cm]{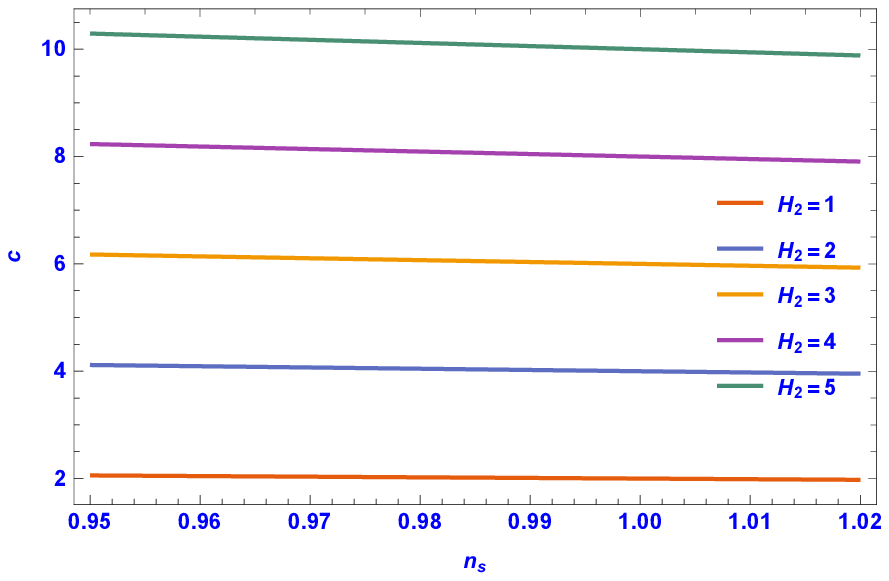}
 \label{2a}}
 \subfigure[]{
 \includegraphics[height=4cm,width=4cm]{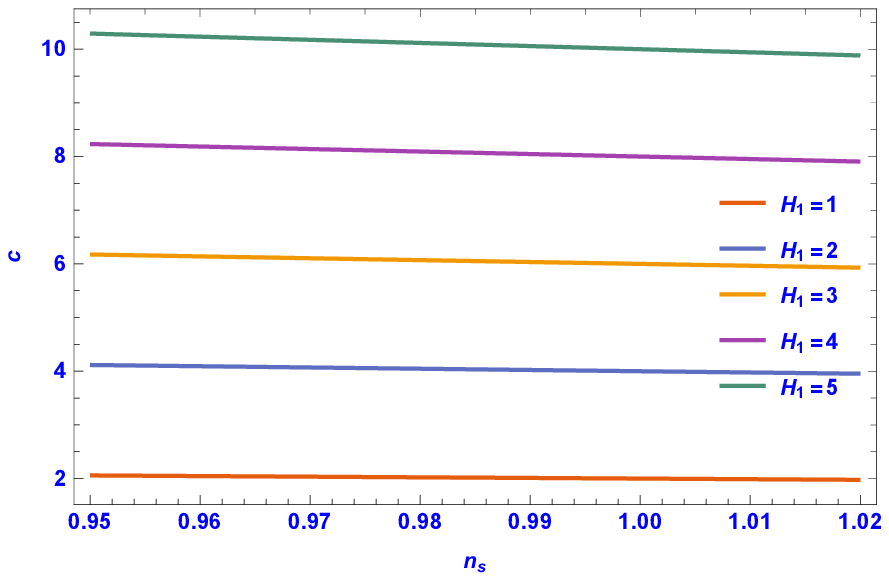}
 \label{2b}}
 \subfigure[]{
 \includegraphics[height=4cm,width=4cm]{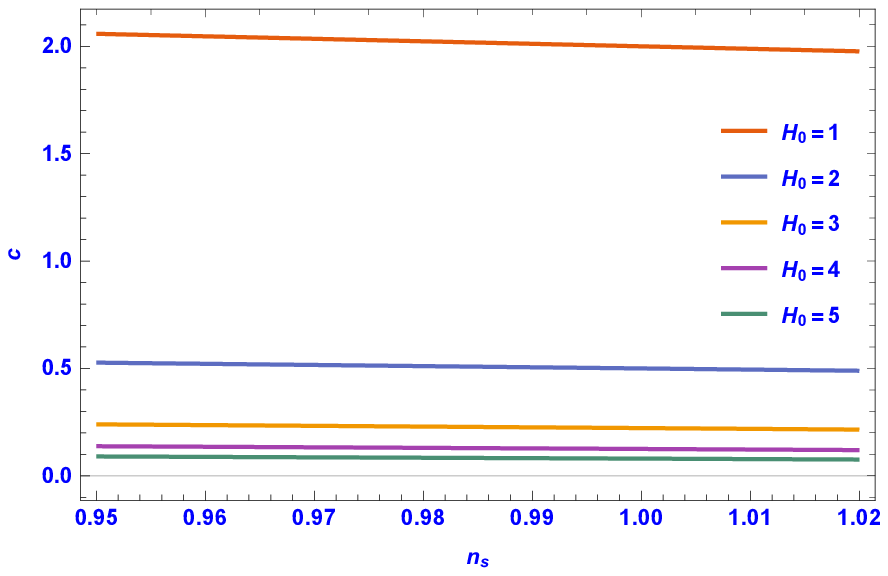}
 \label{2c}}
  \caption{\small{Behavior of the swampland dS conjecture ($c$) in term of $n_{s}$ in units of constant parameters, consistent to the Eq. (\ref{24}).}}
 \label{2}
 \end{center}
 \end{figure}

As shown in the figures, we plotted the values associated with the swampland dS conjecture and the different values obtained for each of the cosmological parameters, such as the scalar spectral index $n_{s}$ in Fig. \ref{1} and Fig. \ref{2} from equations (\ref{23}) and (\ref{24}) as well as the tensor-to-scalar ratio ($r$) in Fig. \ref{3} and Fig. \ref{4} consistent with the change to any of the Hubble parameters. In each plot, except for each mentioned Hubble parameter's changes, we assume other parameters such as $M_{pl}$ as a unit constant positive value. The range associated with these parameters is well determined. As you can see in Fig. \ref{1} and Fig. \ref{2}, the swampland conjecture is well behaved concerning the scalar-spectrum-index for the different values of the Hubble parameter in Fig. \ref{1}, which is derived from the equation (\ref{23}), and shows acceptable values. Similarly, swampland conjectures regarding the tensor-to-scalar ratio as well as the various values of the Hubble parameter per equation (\ref{26}) in Fig. \ref{4} is well defined and shows more acceptable values than by Fig. \ref{3}.\\

\begin{figure}[h!]
 \begin{center}
 \subfigure[]{
 \includegraphics[height=4cm,width=4cm]{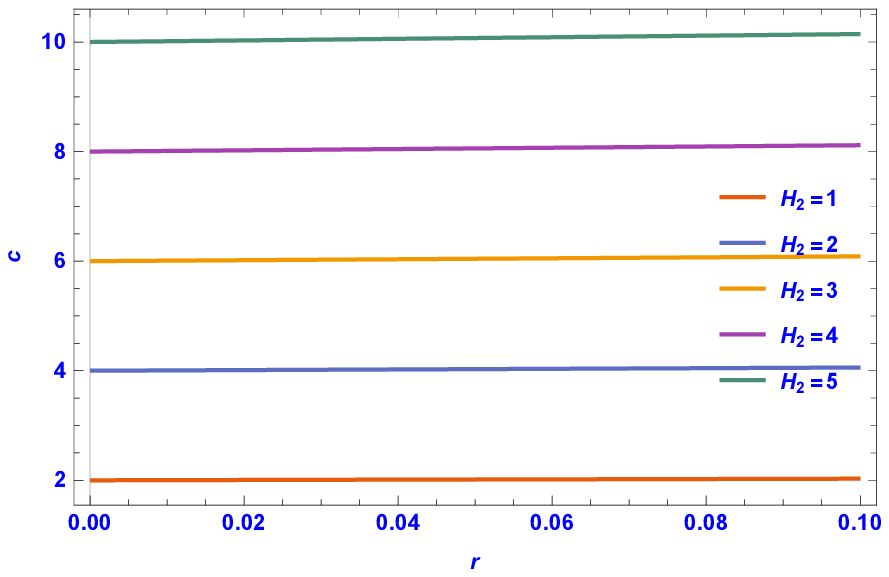}
 \label{3a}}
 \subfigure[]{
 \includegraphics[height=4cm,width=4cm]{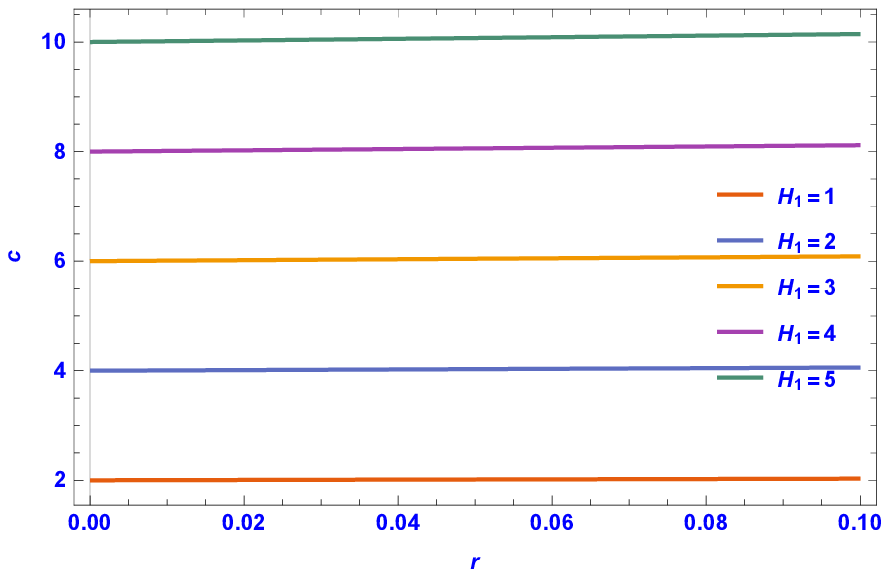}
 \label{3b}}
 \subfigure[]{
 \includegraphics[height=4cm,width=4cm]{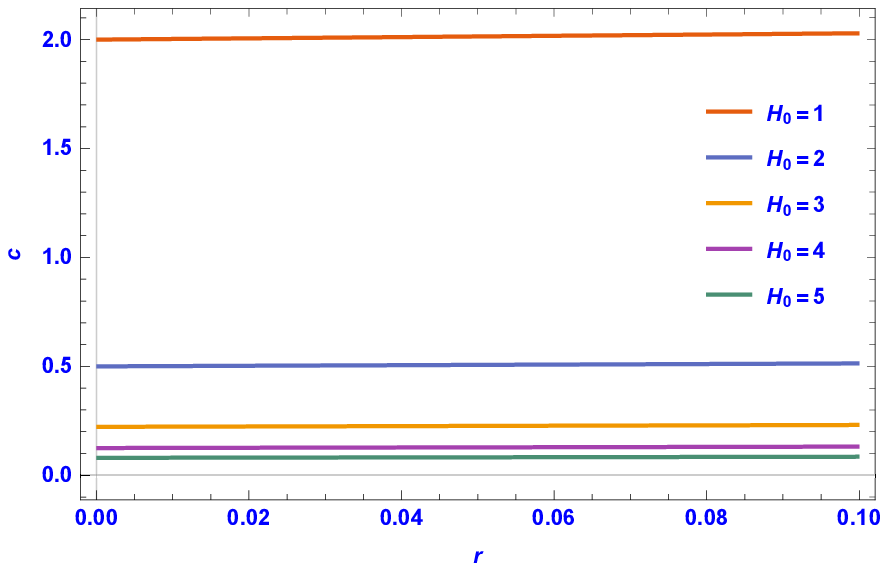}
 \label{3c}}
  \caption{\small{Behavior of the swampland dS conjecture ($c$) in term of tensor-to-scalar ratio $r$ in units of constant parameters, consistent to the Eq. (\ref{25}).}}
 \label{3}
 \end{center}
 \end{figure}

It is clear from Fig. \ref{4} that $c$ parameter is increasing function of $r$ as well as Hubble parameters.\\

\begin{figure}[h!]
 \begin{center}
 \subfigure[]{
 \includegraphics[height=4cm,width=4cm]{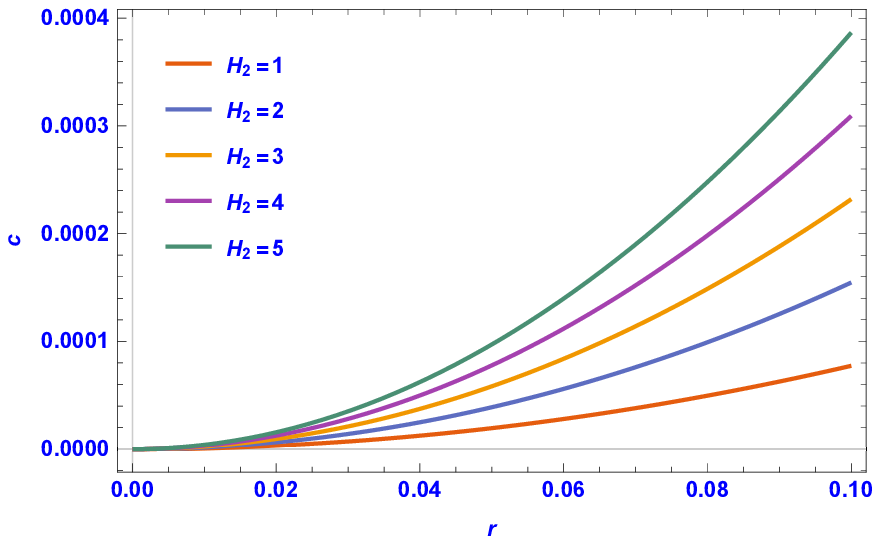}
 \label{4a}}
 \subfigure[]{
 \includegraphics[height=4cm,width=4cm]{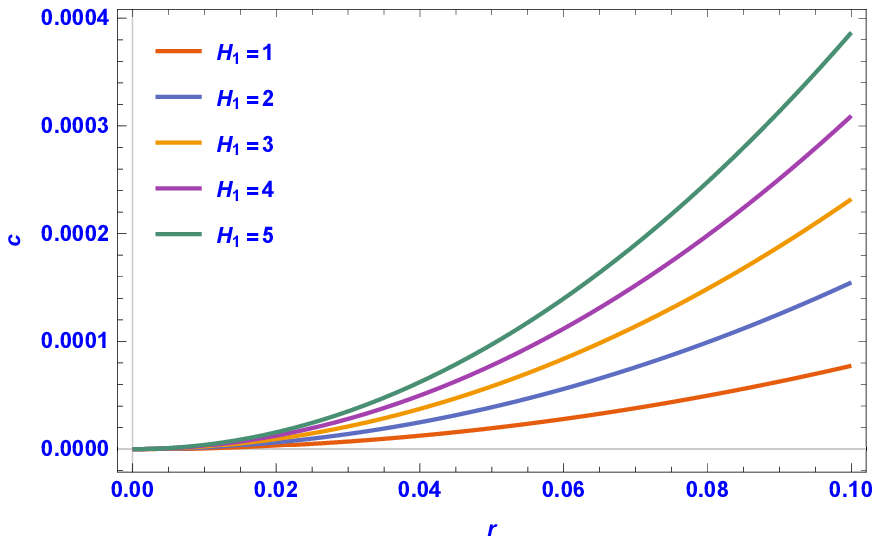}
 \label{4b}}
 \subfigure[]{
 \includegraphics[height=4cm,width=4cm]{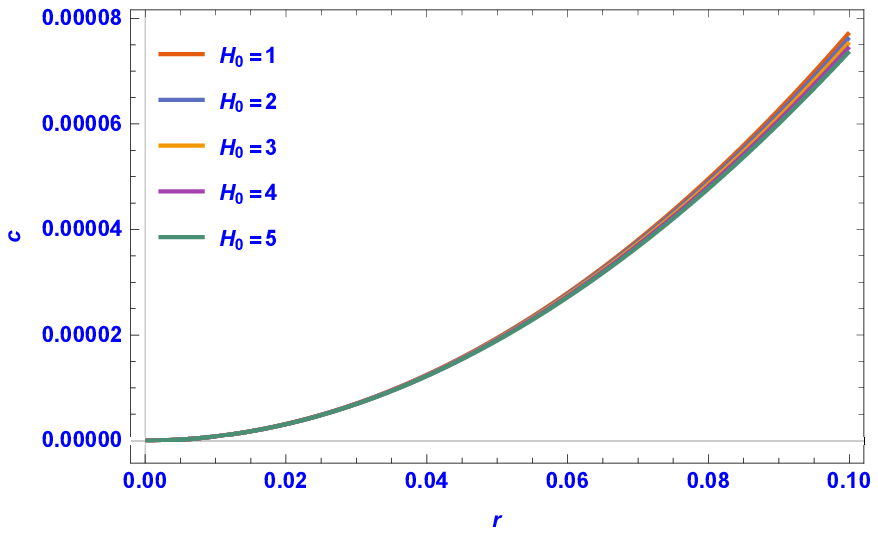}
 \label{4c}}
  \caption{\small{Behavior of the swampland dS conjecture ($c$) in term of tensor-to-scalar ratio $r$ in units of constant parameters, consistent to the Eq. (\ref{26}).}}
 \label{4}
 \end{center}
 \end{figure}

Next, by using the equations (\ref{21}) and (\ref{22}), we plot the cosmic parameters $r$ in terms of $n_{s}$ by Fig. \ref{5}. The range of each of these parameters is determined. As you can see in the Fig. \ref{5}, these values are comparable to observable data. In these figures, we examined the change of the two cosmological parameters of the scalar-spectrum-index $(n_{s})$ and the tensor-to-scalar ratio $r$ for differences of the Hubble parameters. We assumed that the other values were positive of a unit order. It is interesting to note that the range and values obtained for single-field inflation models \cite{das,86} are more accurate than the two-field model according to the Swampland conditions and are consistent with observable data.\\

\begin{figure}[h!]
 \begin{center}
 \subfigure[]{
 \includegraphics[height=4cm,width=4cm]{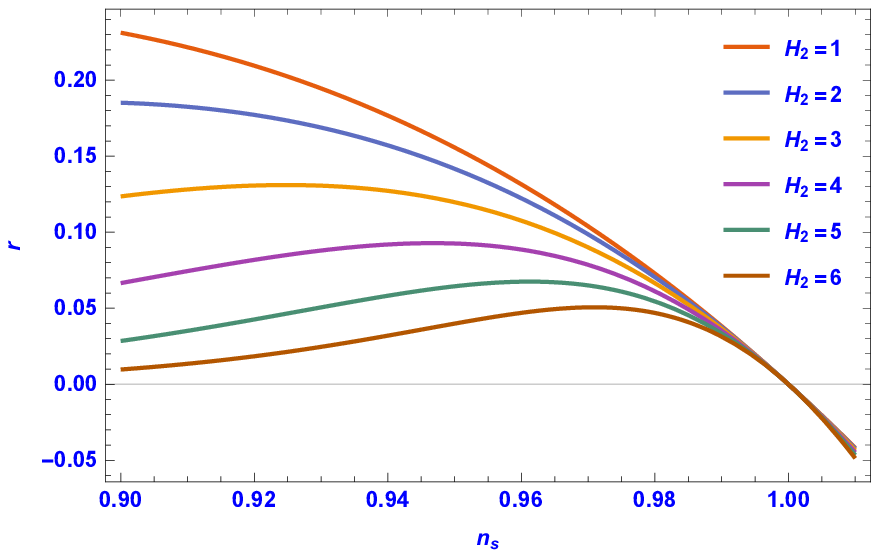}
 \label{6a}}
 \subfigure[]{
 \includegraphics[height=4cm,width=4cm]{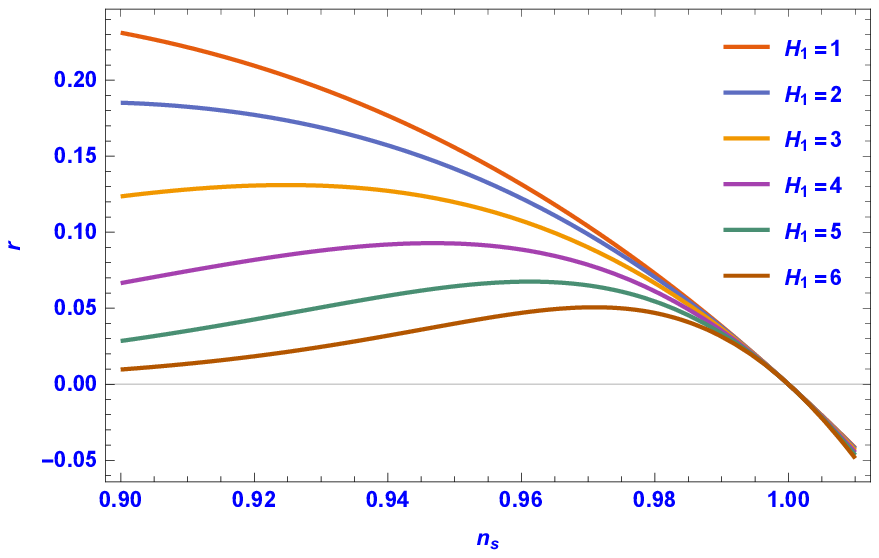}
 \label{6b}}
 \subfigure[]{
 \includegraphics[height=4cm,width=4cm]{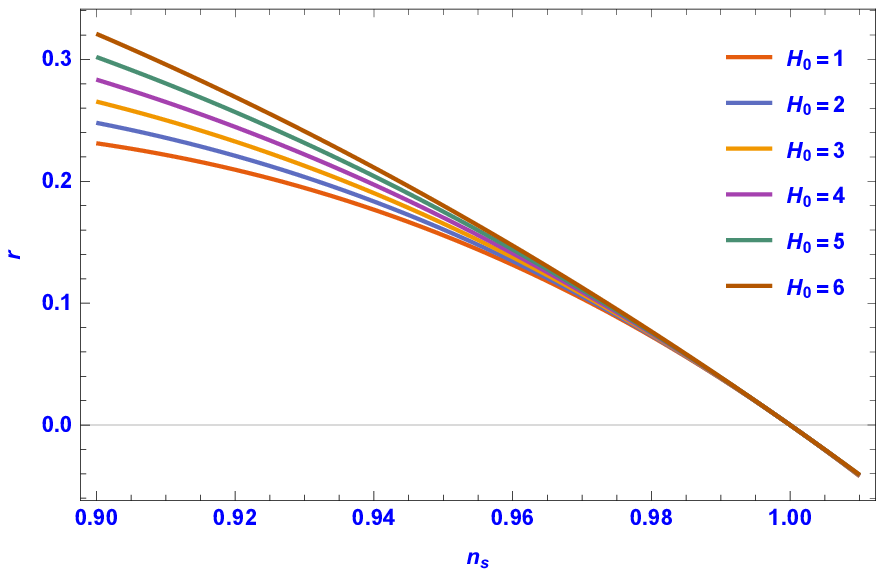}
 \label{6c}}
  \caption{\small{The $(r-n_{s})$ plan in units of constant parameters. (a) by variation of $H_{2}$; (b) by variation of $H_{1}$; and (c) by variation of $H_{0}$.}}
 \label{6}
 \end{center}
 \end{figure}

The allowable area for the scalar-spectral-index and the tensor-to-scalar ratio determined according to the Hubble parameter. As mentioned above, the obtained range is somewhat comparable to the observable data \cite{77}. Here this model compared to single-field inflation \cite{86}  with the slow-roll condition concerning swampland conjecture, it is less accurate. So,  the single-field inflation with swampland conjecture can be more critical in cosmological studies. This paper introduces a two-field inflation model, and we obtained each of the cosmological parameters separately with analytical and numerical analysis. According to the swampland dS conjecture, we examined the inflation model. We determined the range associated with each of the cosmological parameters by plotting some figures consistent with the observable data. In the next works, we will evaluate other implications of this cosmological model according to different conditions.

\section{Conclusions}
Different inflation models according to various conditions such as slow-roll, ultra-slow-roll, constant-roll, and other conditions already investigated in literatures, where many cosmological implications evaluated. In this paper, we investigated a new perspective of the two-field inflation model with respect to the new swampland dS conjecture. Therefore, we studied the two-field inflation model and some cosmological parameters. Then, we obtained the mentioned parameter such as the scalar spectral index $n_{s}$ and tensor-to-scalar ratio ($r$). We applied the mentioned swampland dS conjecture in this paper to our model concerning the observational data like Planck 2018 \cite{1807.06209}.\\
Finally, we compared these values with the observable data by plotting some figures. As mentioned above, the obtained range is somewhat comparable to the observable data. Here, this model compared to single-field inflation with the slow-roll condition concerning swampland conjecture, it is less accurate. So,  the single-field inflation with swampland conjecture can be more important in cosmological studies. In general, the concepts expressed in this article can be examined for the multi-field inflation model by using other ideas such as slow-roll and constant-roll conditions. Also, it is interesting to apply two-field scenario to the M-theory compactifications \cite{88}. In the future works, we will discuss these concepts.

\end{document}